\documentclass[conference]{IEEEtran}
\usepackage{cite}
\usepackage{amsmath,amssymb,amsfonts}
\usepackage{algorithmic}
\usepackage{graphicx}
\usepackage{textcomp}
\usepackage{xcolor}
\usepackage{booktabs}
\usepackage{subfigure}
\usepackage{colortbl}
\newcommand{\ra}[1]{\renewcommand{\arraystretch}{#1}}
\def\BibTeX{{\rm B\kern-.05em{\sc i\kern-.025em b}\kern-.08em
    T\kern-.1667em\lower.7ex\hbox{E}\kern-.125emX}}
\usepackage{enumitem}
\usepackage{multirow}
\usepackage{colortbl}
\usepackage{booktabs}
\usepackage{hhline}    
\usepackage[T1]{fontenc}
\usepackage{tikz}
\usepackage[most]{tcolorbox}
\newtcolorbox{Summary}{
    sharpish corners, 
    boxrule = 0pt,
    toprule = 3.5pt, 
    enhanced,
    fuzzy shadow = {0pt}{-2pt}{-0.5pt}{0.5pt}{black!35} 
}
\definecolor{myblue}{HTML}{1F618D}
\newcommand{\method}{\textit{EvoClass}}

\begin{document}

\title{Cost Reduction on Testing Evolving Cancer Registry System
}

\author{%
\IEEEauthorblockN{%
Erblin~Isaku\IEEEauthorrefmark{1}\IEEEauthorrefmark{2},
Hassan~Sartaj\IEEEauthorrefmark{1},
Christoph~Laaber\IEEEauthorrefmark{1},
Tao~Yue\IEEEauthorrefmark{1},
Shaukat~Ali\IEEEauthorrefmark{1}\IEEEauthorrefmark{3},
Thomas~Schwitalla\IEEEauthorrefmark{4},
and
Jan~F.~Nygård\IEEEauthorrefmark{4}\IEEEauthorrefmark{5}}
\IEEEauthorblockA{\IEEEauthorrefmark{1}Simula Research Laboratory, Oslo, Norway\\
\{erblin, hassan, laaber, tao, shaukat\}@simula.no}
\IEEEauthorblockA{\IEEEauthorrefmark{2}University of Oslo,
Oslo, Norway}
\IEEEauthorblockA{\IEEEauthorrefmark{3}Oslo Metropolitan University, Oslo, Norway}
\IEEEauthorblockA{\IEEEauthorrefmark{4}Cancer Registry of Norway, Oslo, Norway\\
\{thsc, jfn\}@kreftregisteret.no}
\IEEEauthorblockA{\IEEEauthorrefmark{5}UiT The Arctic University of Norway, Tromsø, Norway}}

\maketitle
\thispagestyle{plain}
\pagestyle{plain}

\begin{abstract}
The Cancer Registration Support System (CaReSS), built by the Cancer Registry of Norway (CRN), is a complex real-world socio-technical software system that undergoes continuous evolution in its implementation. Consequently, continuous testing of CaReSS with automated testing tools is needed such that its dependability is always ensured. Towards automated testing of a key software subsystem of CaReSS, i.e., GURI, we present a real-world application of an extension to the open-source tool EvoMaster, which automatically generates test cases with evolutionary algorithms. We named the extension {\method}, which enhances EvoMaster with a machine learning classifier to reduce the overall testing cost. This is imperative since testing with EvoMaster involves sending many requests to GURI deployed in different environments, including the production environment, whose performance and functionality could potentially be affected by many requests. The machine learning classifier of {\method} can predict whether a request generated by EvoMaster will be executed successfully or not; if not, the classifier filters out such requests, consequently reducing the number of requests to be executed on GURI. We evaluated {\method} on ten GURI versions over four years in three environments: development, testing, and production. Results showed that {\method} can significantly reduce the testing cost of evolving GURI without reducing testing effectiveness (measured as rule coverage) across all three environments, as compared to the default EvoMaster. Overall, {\method} achieved $\approx$31\% of overall cost reduction. Finally, we report our experiences and lessons learned that are equally valuable for researchers and practitioners.   

\end{abstract}

\begin{IEEEkeywords}
Software Evolution, Testing, Machine Learning 
\end{IEEEkeywords}

\section{Introduction}
Mandated by the Norwegian government, the Cancer Registry of Norway (CRN) gathers data about all cancer types occurring in the Norwegian population and performs tasks such as producing statistics for policymakers and supporting research by providing relevant data to researchers and other stakeholders. Key functionalities of these tasks are supported by a socio-technical software system named Cancer Registration Support System (CaReSS). Naturally, CaReSS experiences continuous evolution due to many reasons, including software updates, changes in legislation, and new medical standards emerging related to cancers. As a result, CaReSS shall be tested continuously with automated testing tools to ensure that, at any given time, it doesn't produce incorrect data and statistics.

To perform cost-effective testing of evolving CaReSS, we present our real-world application together with experiences of testing ten different versions of a key component of CaReSS-- called GURI. The GURI software system collects and aggregates heterogeneous data coming to CaReSS, e.g., from hospitals and labs. Next, GURI performs validation and aggregation on data with implemented rules that constantly change. Our main objective was to reduce the overall cost of testing GURI by reducing the number of requests that a testing tool needs to make to GURI, while at the same time not compromising the testing effectiveness, measured as rule coverage in our context. 

In this work, we rely on a well-known, open-source, AI-enabled, and system-level testing tool called EvoMaster~\cite{arcuri2019restful}. EvoMaster automates testing through Representational State Transfer (REST) Application Programming Interfaces (APIs) with search algorithms. Since GURI exposes REST APIs, it is natural for us to select a tool that can automate testing through REST APIs. Since EvoMaster generates test cases with many requests to GURI through REST APIs, significantly increasing interactions with GURI, which thereby incurs significant costs on test execution and potentially impacts the performance of GURI. To reduce the costs incurred by many requests, we train a machine learning (ML) classifier that can predict whether a particular request is likely to fail during its execution; if so, the classifier rejects such requests. As a result, EvoMaster empowered with the ML classifier can focus on successful requests. This extension to EvoMaster is named {\method}. 

To assess the cost-effectiveness of {\method}, we tested ten GURI versions, which were naturally formed over four years of its evolution under three environments, i.e., development, testing, and production. Results show that {\method} can significantly reduce testing cost ($\approx$31\%) compared to the default EvoMaster, while not reducing testing effectiveness in terms of rule coverage. Based on testing GURI, we provide a set of lessons learned that are valuable for practitioners and researchers focusing on testing similar kinds of software systems. 

We organize the paper as follows. The background is given in Section~\ref{sec:background}; the proposed approach is described in Section~\ref{approach}; empirical evaluation is presented in Section~\ref{sec:evaluation}; experiences and lessons learned are discussed in Section~\ref{sec:lessons}; the related work section is described in Section~\ref{relatedworks}; and the paper concludes in Section~\ref{conclusion}.

\section{Background} \label{sec:background}

\subsection{Real-World Context and Challenges}\label{background}
The Cancer Registry of Norway (CRN) is a public organization gathering data and statistics about cancer patients, e.g., diagnostic details, treatment records, and follow-up information. 
Such data are made available to various end users, including researchers, patients, doctors, and healthcare authorities. CRN has developed an interactive decision support system named Cancer Registration Support System (CaReSS) to ensure the accuracy of the data and statistics being released to end users~\cite{laaber2023challenges}. CaReSS, as a rule-based system, undergoes continuous evolution as rules (e.g., a cancer diagnosis date cannot be before the patient's birth date) are added, modified, or removed due to new treatments, improved diagnostics, advances in medical findings, and updated diagnostic standards~\cite{wang2017rcia}. One component of CaReSS, named GURI, automatically validates and aggregates collected data against the rules implemented in it, defined by domain experts. Patients' data (i.e., cancer messages) are sent to GURI via a web application through REST APIs. Specifically, we focus on two REST endpoints: the validation endpoint, which validates cancer messages against validation rules, and the aggregation endpoint, responsible for consolidating cancer messages into cancer cases. 

Testing GURI is crucial because incorrect or imprecise statistics can significantly influence research findings and decisions made by relevant stakeholders such as healthcare professionals and authorities~\cite{laaber2023challenges}. 
One important goal of CRN is to enable automated, rigorous, and cost-effective testing of GURI. 
However, the continuous evolution of GURI, especially its implemented rule set, makes its testing very challenging. Moreover, each addition, deletion, or modification of a rule in a particular version of GURI is addressed in multiple environments. Initially, these rules are created in GURI's \textit{development} environment. 
Next, they are moved to the \textit{test} environment for testing them through CaReSS. 
Finally, after corrections (if any), these rules are moved to the \textit{production} environment as a part of CaReSS. Testing each version of GURI in the three different environments using an automated testing technique is costly. Thus, this work aims to reduce the effort in testing GURI in multiple environments and for different versions using ML techniques.


\subsection{System Level Testing with EvoMaster} \label{subsec:EvoMaster}
EvoMaster~\cite{arcuri2019restful} is an open-source tool that uses evolutionary algorithms for supporting system-level testing of enterprise-level web APIs developed using REST and GraphQL. It has been continuously developed for over six years with the addition of new functionalities~\cite{arcuri2023building}. It takes web APIs' schema in OpenAPI specification (OAS) or Swagger format and generates test cases in Java, Kotlin, and C\#. 
EvoMaster generates test cases considering multiple objectives, such as fault detection and code coverage (for white-box testing). It also supports black-box testing that uses random and multi-objective search algorithms to optimize various objectives, e.g., success status codes (2XX).


\section{Approach}\label{approach}

We propose {\method}, an ML-based approach to enhance EvoMaster for reducing test execution cost in the context of testing GURI. The underlying idea is to predict the success or failure of RESTful API requests generated by EvoMaster during test generation without actually executing them. With such prediction, our approach discards requests that are highly likely to lead to a desired status code (success or failure), which is not of our interest (i.e., failure in the context of this paper), thereby leading to reduced test execution cost.


Specifically, we trained and integrated Random Forest--an ML classifier (selected with an empirical evaluation, details in Section \ref{sec:evaluation}), to classify the generated test cases. Using the trained ML model, {\method} only executes requests predicted to be executed successfully (i.e., 200 status codes). This approach ensures meaningful responses from HTTP requests, focusing on testing the core functionality, i.e., validating cancer messages with validation rules and aggregating cancer messages into cancer cases with aggregation rules in our healthcare-specific domain rather than just input validation. 

\begin{figure*}[htbp]
  \centering

\includegraphics[width=0.8\textwidth]{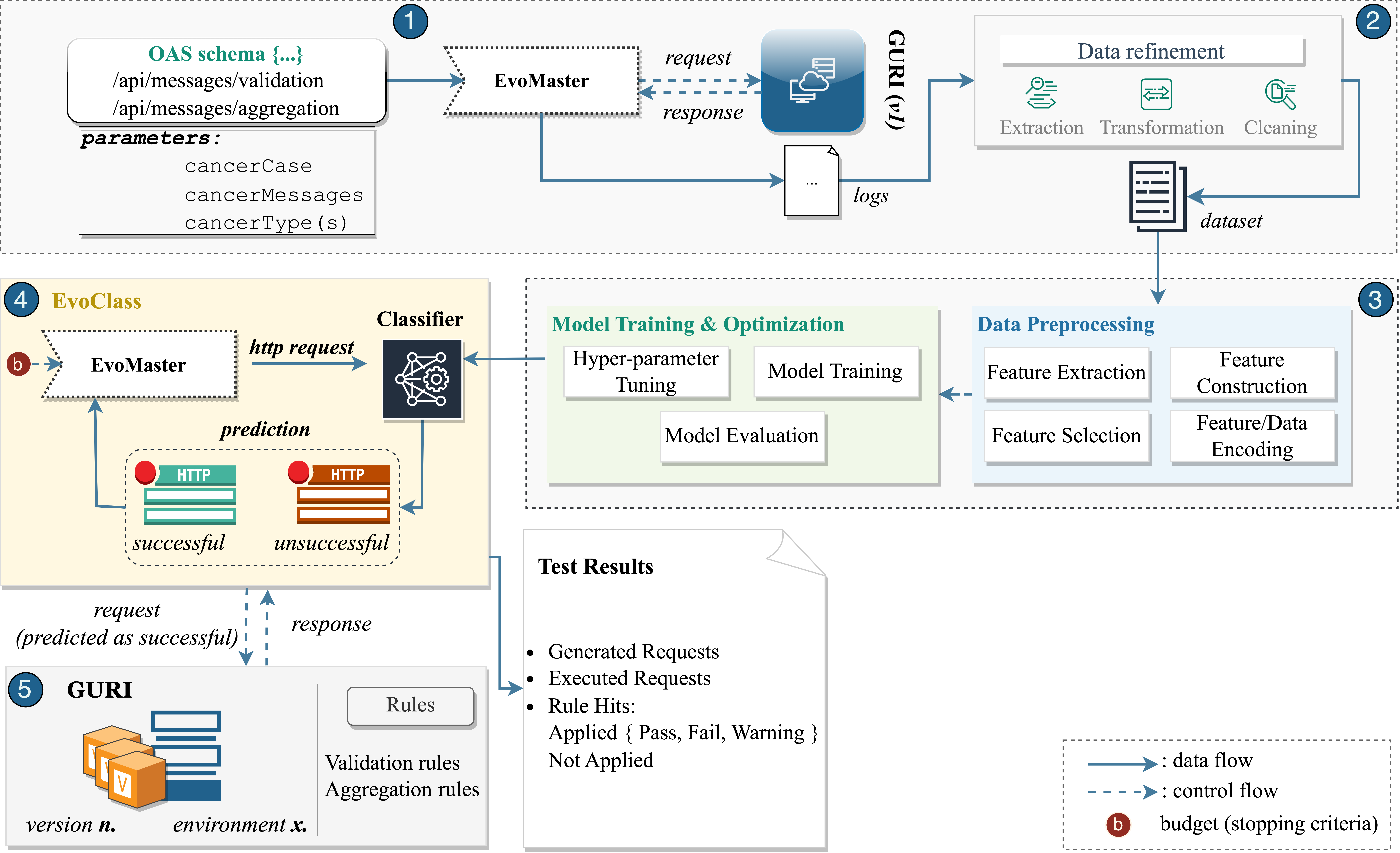}
  \caption{Overview of {\method}}
    \label{fig:approach}
\end{figure*}

Figure~\ref{fig:approach} shows our approach's four components: data collection (Section \ref{subsec:datacollection}), data preprocessing (Section \ref{subsec:datapreprocessing}), model training and optimization (Section \ref{subsec:training and optimization}), and integration with EvoMaster (Section \ref{subsec:integration}). 
In the data collection phase \begin{tikzpicture}
  \filldraw[fill=myblue] (0,0) circle (0.2cm);
  \node[text=white, scale=0.7] at (0,0) {\textbf{1}};
\end{tikzpicture}, we use EvoMaster as a REST API testing tool to generate test cases automatically. 
The two API endpoints that undergo testing are related to the validation and aggregation of medical rules implemented in GURI and log all the data related to requests (e.g., generated inputs) and their respective responses (e.g., status code). The collected data is then refined \begin{tikzpicture}
  \filldraw[fill=myblue] (0,0) circle (0.2cm);
  \node[text=white, scale=0.7] at (0,0) {\textbf{2}};
\end{tikzpicture} through the extraction, transformation, and cleaning steps to prepare it for further analysis and preprocessing. Data preprocessing \begin{tikzpicture}
  \filldraw[fill=myblue] (0,0) circle (0.2cm);
  \node[text=white, scale=0.7] at (0,0) {\textbf{3}};
\end{tikzpicture} involves techniques such as feature extraction, construction, selection, and encoding to prepare the dataset for classification. The next step is about model training and optimization, where we split the dataset, use the Random Forest classifier as our model, and optimize its hyperparameters using the Optuna framework. Finally, we integrate the trained model into EvoMaster \begin{tikzpicture}
  \filldraw[fill=myblue] (0,0) circle (0.2cm);
  \node[text=white, scale=0.7] at (0,0) {\textbf{4}};
\end{tikzpicture}, enabling real-time prediction and selective execution of requests based on their predicted success \begin{tikzpicture}
  \filldraw[fill=myblue] (0,0) circle (0.2cm);
  \node[text=white, scale=0.7] at (0,0) {\textbf{5}};
\end{tikzpicture}.


\subsection{Data Collection} \label{subsec:datacollection}
We use EvoMaster to generate test data, i.e., requests, to train the ML model. For training, we categorize generated requests based on the response status code retrieved from the HTTP calls by following the RFC 9110 standard~\cite{rfc9110}. 
Specifically, a request is successful if the status code is 200, representing an "OK" response from the server. On the other hand, requests are categorized as failures if the status code differs from 200. Typically, requests with 4XX status codes (client-side errors) and 5XX status codes (server-side errors) are considered failures. However, in our specific case, we have not encountered any requests resulting in a 4XX status code. Instead, we have identified cases where a 302 status code is returned, which we have determined to be related to the authorization process. Our investigation also revealed that EvoMaster triggers the 302 status code response when it executes requests without authorization headers. This is also categorized as an unsuccessful call, i.e., a failure.

In this specific scenario (i.e., 302 status code outcomes), manually adjusting some filtering conditions to exclude these test cases can be effective. However, it is not straightforward in cases of 4XX and 5XX status codes due to request body parameters. For example, filtering such cases requires identifying invalid/malformed requests dependent on query, path, and body parameters/content.
Therefore, the complexity of identifying all possible invalid combinations/patterns leads to employing ML in our approach. 

\begin{figure}[htbp]
  \centering

\includegraphics[width=0.4\textwidth]{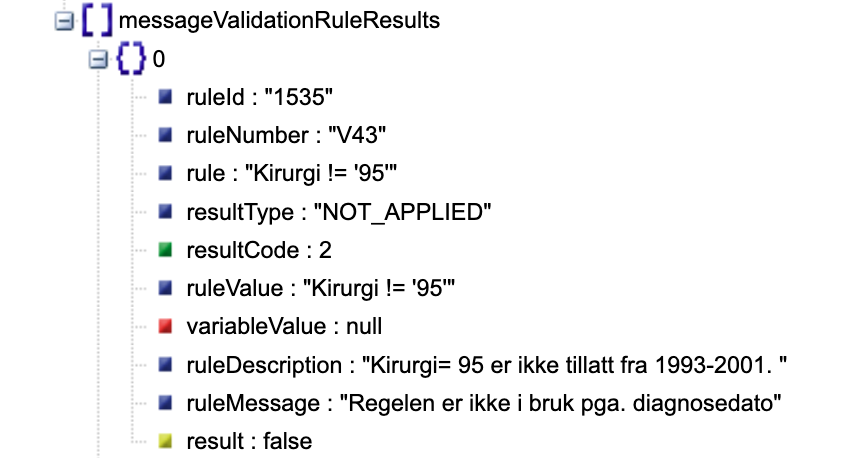}
  \caption{Response snippet of a successful request}
    \label{fig:sucessful-req}
\end{figure}

We used the following two API endpoints of CaReSS that are relevant for GURI for input generation based on OpenAPI Specification (OAS), Swagger \cite{swagger}. The request method of both endpoints is \textit{POST}: \textit{a). /api/messages/validation, b). /api/messages/aggregation}.
EvoMaster automatically generates valid and/or invalid inputs for the selected endpoints based on the respective OAS schema. A valid input is defined based on the expected requirements of each endpoint, such as data types (e.g., string) and different constraints (e.g., date format). After successfully creating the request consisting of the endpoint, method, and body parameters, EvoMaster executes the request and returns a response (including the status code). We aim to retrieve \textit{meaningful} responses, which occur in successful requests (with 200 status code responses), as shown in Figure~\ref{fig:sucessful-req}.

We slightly adapted EvoMaster to log each request/response after each execution to create a training dataset for the ML model. To obtain raw data, EvoMaster was running for a continuous duration of 10 hours in the initial stable version (v1) of GURI. During this runtime, EvoMaster is unaware of the difference in the environments (i.e., test, development, and production) in which GURI is deployed. This is important to ensure that we learn an ML model that can make predictions in any given environment. 

This comprehensive log file is then refined and prepared in a suitable format for training the model. 
The dataset refinement and preparation involve three main steps: data extraction, transformation, and cleaning:
    
    \noindent\textbf{{Data extraction.}} This step parses objects and strings to extract relevant information, e.g., the API URL and authorization status.
    
    \noindent\textbf{Transformation.} This step organizes the extracted data structurally by converting nested objects into dictionaries.
    
    \noindent\textbf{Cleaning.} This step removes irrelevant or redundant information. For instance, the raw data included detailed information that was only accessible after executing a request, such as fitness scores or covered targets (EvoMaster-related metrics). Since we aim to predict a request before executing, these metrics would not be beneficial as they are not present during the request generation. 
%
The final dataset contains 13,985 records (including status codes). Table \ref{tab:dataset_distribution} shows the distribution of the records in the dataset in terms of status codes for both the validation rule and aggregation rule endpoints.
\begin{table}[htbp]
\caption{Distribution of the records in the dataset across three status code types and two endpoint types}
\centering
\begin{tabular}{rrr}
\toprule
\textbf{Status Code} & \multicolumn{2}{c}{\textbf{Rule Endpoint}} \\
\cmidrule{2-3}
& \textbf{Aggregation} & \textbf{Validation} \\
\midrule
200              & 26.73\%               & 33.32\%             \\
\rowcolor[HTML]{f9f9f9}
500              & 20.51\%                         & 14.65\%         \\  
302              & 2.31\%                         & 2.48\%         \\  \midrule
\textbf{Total} & 49.55\% & 50.45\%\\

\bottomrule
\end{tabular}
\label{tab:dataset_distribution}
\end{table}


\subsection{Data Preprocessing} \label{subsec:datapreprocessing}
We applied several preprocessing techniques to ensure the data's consistency and suitability for training the ML classifier, which are described below: 

\subsubsection{Feature extraction} This step selects relevant information from the data and represents it in a way that captures important patterns or characteristics.
   
    \noindent \textbf{Converting the target variable:} We convert the original target variable (i.e., status code), into a binary classification problem. We assign a value of "1" to successful requests (i.e., a 200 status code) and "0" to all other cases (i.e., 302 and 500 status codes). As a result, the target variable is in a binary format, thereby suitable for binary classification.
    
    \noindent \textbf{Decomposing categorical attributes:} We decompose categorical variables into binary features. For example, in our case, we have a feature related to authentication (Auth). The "Auth" feature is decomposed into a new binary feature called {\fontfamily{qcr}\selectfont"is\_no\_auth"}. This new feature takes the value "1" if the authentication is missing and "0" otherwise. As a result, we expose more information to the model in a format that is easier to process.
    
    \noindent \textbf{Decomposing a date-time:} Date-time attributes are rich in structure and have relationships with other attributes. In this case, the date-time attribute, such as {\fontfamily{qcr}\selectfont"cancerCase.diagnosedato"}, is decomposed into binary features using regular expressions. These binary features indicate whether the date has a valid format or not. A value of "1" is assigned if the format is valid, and "0" otherwise. This decomposition simplifies the attribute and allows the model to focus on the validity aspect rather than the specific date-time values.

\subsubsection{Feature construction} We created {\fontfamily{qcr}\selectfont"cancerMessagesNr"} and {\fontfamily{qcr}\selectfont"cancerTypesNr"} as new features to quantify the number of cancer messages and their types. By counting the occurrences of cancer messages and their types, these new features provide additional information about the presence or frequency of cancer-related data to assist the model in recognizing patterns between cancer types and messages that can lead to faulty requests.

\subsubsection{Feature selection} To determine the relevance of each feature in our dataset, we utilized impurity-based feature importance \cite{breiman2001random} from scikit-learn \cite{scikit-learn}. We wanted to evaluate the importance of the newly created features. Thus, we computed and examined the feature importance scores during model training. 
The feature selection process is performed iteratively within the training phase, meaning that an irrelevant feature will be dropped, and model training will continue with the updated feature set. Interestingly, we found that the {\fontfamily{qcr}\selectfont"Method Type"} feature had an importance score of 0, indicating that it is irrelevant to our task. Therefore, the "Method Type" feature was excluded from the dataset. This step allowed us to streamline our feature set and focus on the more informative features of our classifier model.

\subsubsection{Feature/Data Encoding} 

In this step, we applied {\fontfamily{qcr}\selectfont label encoding} from scikit-learn \cite{scikit-learn} to transform certain \textit{qualitative} input variables \textit{(e.g., user, cancer type, and environment)} into numerical representations. This ensures that our model can effectively process and interpret the data.

We use label encoding instead of other techniques, e.g., one-hot encoding, due to the specific characteristic of our dataset, i.e., the predominance of qualitative variables. 
Another reason for using label encoding is the feasibility of model evaluation since we can avoid the issue of missing features during model evaluation. For instance, variable \textit{"Cancer Type"} represents different types of cancers, such as \textit{"Breast Cancer," "Lung Cancer," "Prostate Cancer," and "Colon Cancer"}. By using label encoding, we assign numerical labels to each category (e.g., 0 for Breast Cancer, 1 for Lung Cancer, 2 for Prostate Cancer, and 3 for Colon Cancer) instead of creating a new feature (in the case of one-hot encoding).


\subsection{Training and Optimization} \label{subsec:training and optimization}

This step trains and optimizes the ML model. The dataset, represented by the feature matrix X and target variable Y, was split into training and testing sets of 80\% and 20\%, respectively. We employed the {\fontfamily{qcr}\selectfont Random Forest Classifier} as our model, selected based on a pilot experiment (results shown in Figure~\ref{fig:performance_plot} and Table~\ref{tab:comparison_report}), and used the scikit-learn library to fit the model to the training data. 

To optimize the hyperparameters, we used the {\fontfamily{qcr}\selectfont Optuna} framework. We defined a set of parameters to search over, including the number of estimators, maximum depth, minimum samples split, minimum samples leaf, and maximum features. Optuna systematically searched these parameter combinations to find the configuration that achieved the highest accuracy.
The optimal parameters determined by Optuna were as follows: {\fontfamily{qcr}\selectfont n\_estimators=100}, {\fontfamily{qcr}\selectfont max\_features=None}, {\fontfamily{qcr}\selectfont min\_samples\_split=2}, {\fontfamily{qcr}\selectfont min\_samples\_leaf=10}, and {\fontfamily{qcr}\selectfont max\_depth=10}.
These optimized parameters were used to train the model.

\subsection{Integration into EvoMaster}\label{subsec:integration}

\begin{figure}[htbp]
  \centering

\includegraphics[width=0.49\textwidth]{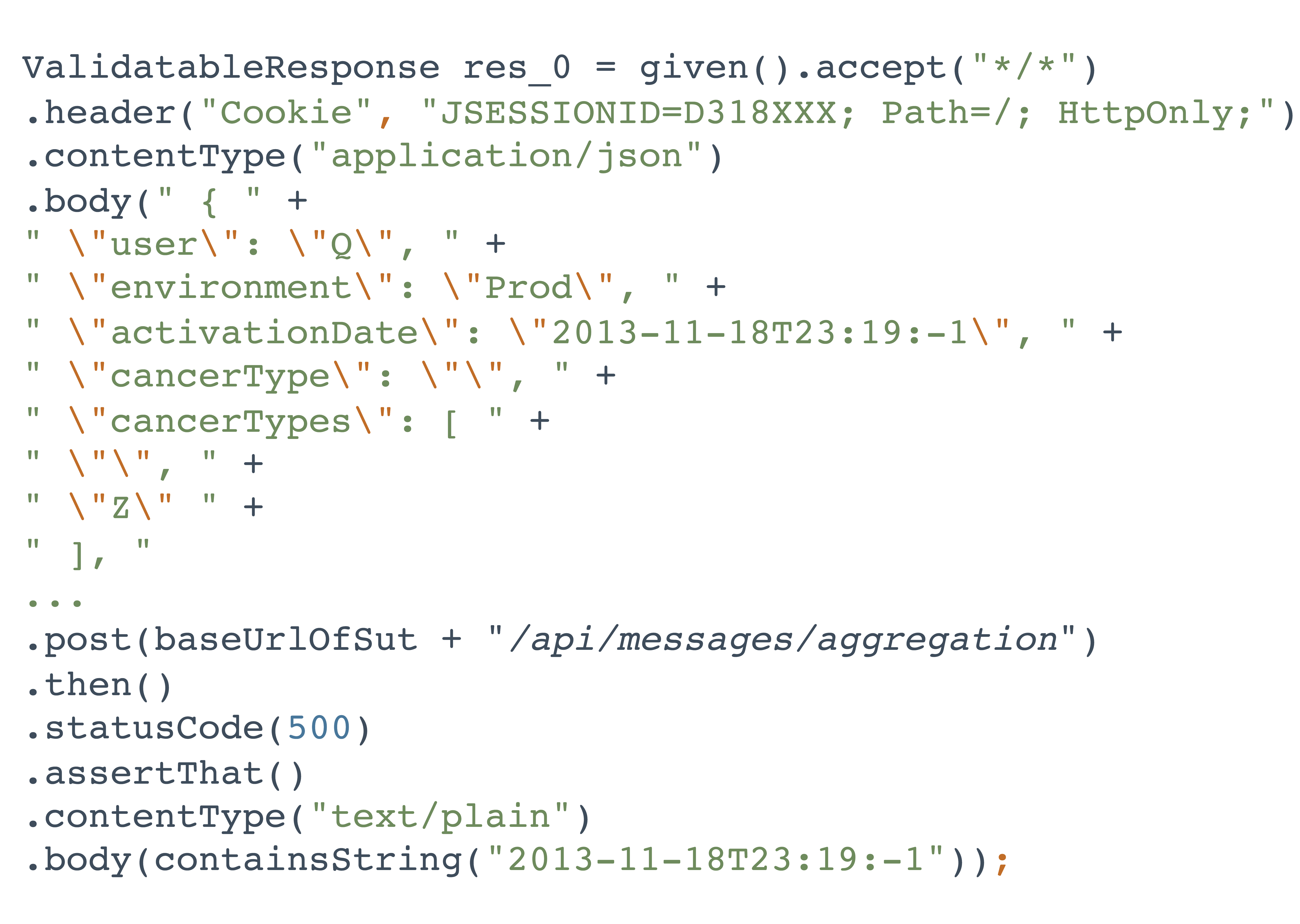}
  \caption{A test case generated by EvoMaster leading to a 500 status code}
    \label{fig:unsuccessful-req}
\end{figure}

The trained ML was serialized and stored using the pickle library \cite{van1995python} followed by integrating it into EvoMaster with the following method. The method communicates with the ML model by sending data requests (e.g., the generated data body) and awaiting the model's prediction. The corresponding request would be executed if the prediction indicated a successful status code. Otherwise, the request is removed from further processing. An example of such requests (i.e., test cases) is shown in Figure~\ref{fig:unsuccessful-req}. The method involves iterating over the actions of an individual request generated by EvoMaster. Each action is formatted into a JSON object and passed as input to a Python script through a subprocess. The subprocess is executed using the {\fontfamily{qcr}\selectfont ProcessBuilder } class, facilitating the interaction between the Java and Python scripts. With this integration, we used the trained model for real-time prediction and decision-making, ensuring the execution of only successful requests, as predicted by the model.

\section{Evaluation}\label{sec:evaluation}
We describe research questions, subject application, evaluation setup, execution, metrics, and discussion. 


\subsection{Research Questions}
\begin{itemize}[leftmargin=10pt]
    \item \textbf{RQ0:} Which ML model performs the best for the classification task of {\method}? We aim to find the most suitable ML model to be integrated into {\method}. We experimented with four commonly used classifiers that can be efficiently trained without requiring a large amount of data: Random Forest, Logistic Regression, KNeighborsClassifier, and GaussianNB. 
        \item \textbf{RQ1:} How effectively does {\method} reduce the testing cost? We study whether the ML model effectively filters out possibly failing requests. Since we have ten GURI versions deployed on three different environments, we also check whether {\method} can obtain consistent performance across the versions and environments. 
    \item \textbf{RQ2:} How much rule coverage is achieved by {\method} compared to the baseline? We aim to know whether {\method} can achieve comparable rule coverage as the default EvoMaster. 

   
\end{itemize}

\subsection{Subject Application}
GURI is the subject application provided by our collaborator CRN (Section \ref{background}). We selected 10 GURI versions running in the development, test, and production environments. GURI has a total of 32 REST APIs corresponding to different functionalities. Only two REST APIs are related to validation and aggregation rules, which we used. 
This setting is in line with our previous work~\cite{laaber2023automated}.
Table~\ref{tab:rules} shows each GURI version's time stamp and the number of validation and aggregation rules. The first version \textit{v1} contains 30 validation and 32 aggregation rules.
After the evolution of the rule set over four years, the recent version \textit{v10} has, in total, 70 validation rules and 43 aggregation rules.

\begin{table}[htbp]
\caption{Descriptive statistics of the rule set's 10 versions~\cite{laaber2023automated}}
\centering
\begin{tabular}{llrr}
\toprule
\textbf{Version} & \textbf{Date} & \textbf{Validation} & \textbf{Aggregation} \\
\midrule
v1 & 12/2017 & 30 & 32 \\
\rowcolor[HTML]{f9f9f9} 
v2 & 05/2018 & 31 & 33 \\
v3 & 02/2019 & 48 & 35 \\
\rowcolor[HTML]{f9f9f9} 
v4 & 08/2019 & 49 & 35 \\
v5 & 11/2019 & 53 & 37 \\
\rowcolor[HTML]{f9f9f9} 
v6 & 09/2020 & 56 & 37 \\
v7 & 11/2020 & 66 & 38 \\
\rowcolor[HTML]{f9f9f9} 
v8 & 04/2021 & 69 & 43 \\
v9 & 01/2022 & 69 & 43 \\
\rowcolor[HTML]{f9f9f9} 
v10 & 01/2022 & 70 & 43 \\
\bottomrule
\end{tabular}

\label{tab:rules}
\end{table}

\begin{table*}
 \ra{1.2}
\caption{Examples of the rule evolution across environments}
\resizebox{\textwidth}{!}{%
\begin{tabular}{lllll}
\toprule
\textbf{Environment} & \textbf{Rule Nr.} & \textbf{Cancer Type} & \textbf{Validation Rule} & \textbf{Modification Type} \\
\midrule

& & & \textit{\textcolor[HTML]{A93226}{(}Topografi -\textgreater{}startswith ('50') \textcolor[HTML]{A93226}{and Ekstralokalisasjon!=   '7777')}} & \\
\multirow{-2}{*}{Test} & \multirow{-2}{*}{R03} & \multirow{-2}{*}{Breast} & \textit{implies Metastase in {[}'0', 'A', 'B', 'C', 'D', '9'{]}} & \\

\rowcolor[HTML]{f9f9f9}
& & & \cellcolor[HTML]{f9f9f9}\textit{Topografi -\textgreater{}startswith ('50')} & \cellcolor[HTML]{FFFFFF} \\
\rowcolor[HTML]{f9f9f9}
\multirow{-2}{*}{Prod} & \multirow{-2}{*}{R03} & \multirow{-2}{*}{Breast} & \cellcolor[HTML]{f9f9f9}\textit{implies Metastase in {[}'0', 'A',   'B', 'C', 'D', '9'{]}} & \cellcolor[HTML]{FFFFFF} \multirow{-4}{*}{\textit{Delete}} \\

 \midrule

& & & \textit{(Meldingstype = 'K' and Topografi \textcolor[HTML]{AF601A}{notIn {[}'481','482','570',   '579'{]}}} &  \\
\multirow{-2}{*}{Dev} & \multirow{-2}{*}{R40} & \multirow{-2}{*}{All} & \textit{and Topografi-\textgreater{}substring(1,2) \textcolor[HTML]{AF601A}{notIn{[}'51', '52','53', '54', '55',   '56', '61'{]})} implies Metastase != '5'} &  \\

\rowcolor[HTML]{f9f9f9}
& & & \textit{(Meldingstype = 'K' and Topografi notIn {[}'481', '482', '488',   '570', '569', '579', '619'{]}} & \cellcolor[HTML]{FFFFFF} \\
\rowcolor[HTML]{f9f9f9}
\multirow{-2}{*}{Prod} & \multirow{-2}{*}{R40} & \multirow{-2}{*}{All} & \textit{and Topografi-\textgreater{}substring(1,2) notIn{[}'51',   '52', '53', '54', '55'{]}) implies Metastase != '5'} & \cellcolor[HTML]{FFFFFF} \multirow{-4}{*}{\textit{Modify}} \\

 \bottomrule 
\end{tabular}%
}
\label{tab:rule_examples}
\end{table*}


The rule evolution occurs in both versions and environments. It includes creating and refining rules in the development environment, testing and modifying them in the test environment, and deploying the validated rules in the production environment. While most modifications are partial (e.g., additional constraints), there are cases where rules are fully deleted or newly introduced. These iterative processes ensure continuous improvement and accuracy of the rule set in GURI. Table~\ref{tab:rule_examples} illustrates that rules commonly undergo various change types (i.e., deletion, modification, and insertion). For rule R03, which applies to validate cancer messages related to breast cancer, a new condition such as {\fontfamily{qcr}\selectfont Ekstalokalisasjon != '7777'} is introduced to the test environment. However, for the same rule, in the production environment, this constraint is removed, implying that any value for variable \textit{Ekstalokalisasjon} received from cancer messages is acceptable (i.e., being validated to be true). Similarly, for the other example (R40), which applies to all cancer types, we can observe modifications across the environments.

\subsection{Evaluation Setup, Execution, and Metrics}

\textbf{Setup.}
We trained the ML model using a dataset collected by running EvoMaster on the first version of GURI for ten hours to collect training data (Section \ref{subsec:datacollection}). 
We use 80\% data for training and 20\% for validation/testing according to the commonly adapted 80-20 split. 
We select the Random Forest classifier for training the ML model based on the results of a pilot experiment (see details in Section~\ref{RQ0}). 


We compare our approach to EvoMaster in black-box mode~\cite{martin2021black} as the baseline, as Laaber~et~al.~\cite{laaber2023automated} showed that, in the context of the CRN and GURI, EvoMaster in black-box mode performs on-par in terms of coverage and fault detection and is superior regarding rule coverage when compared to EvoMaster in white-box mode. 
We configure GURI's ten versions in three different environments, i.e., development, test, and production environment.  
We repeat each configuration of our experiment 30 times which is recommended for experiments with inherent randomness~\cite{arcuri2011practical}. 
We specified one hour time bound for each repetition, which is a common practice~\cite{kim2022automated}. 
The overall computation time required for our experiment is 2 (approaches) * 3 (environments) * 10 (versions) * 30 (repetitions) * 1 (hour) = 1800 hours (75 days) if run sequentially.

\textbf{Execution.}
We executed experiments on a high-performance computing cluster named Experimental Infrastructure for Exploration of Exascale Computing\footnote{https://www.ex3.simula.no/} (eX3) provided by Simula Research Laboratory. 
Our experiment utilized eight nodes of the eX3 cluster running on the Ubuntu operating system. 
All nodes have 2 GB of RAM, 4 TB GB local NVMe scratch storage, and four different types of processors, including 32-core AMD EPYC™ 7601, 64-core AMD EPYC™ 7763, 24-core AMD EPYC™ 7413, and 24-core Intel® Xeon® Platinum 8168.  
The eX3 cluster uses Slurm\footnote{https://slurm.schedmd.com/} for resources management.

\textbf{Metrics and Statistical Tests.}
For RQ0, in addition to using Receiver Operating Characteristics (ROC) curve and Area Under Curve (AUC), we measure accuracy, precision, recall, and F1-score, which are commonly used metrics for evaluating ML model performance. 
To analyze results for RQ1, we also calculate accuracy, precision, recall, and F1-score. In addition, we introduce the metric of cost reduction, which is with the formula below:
\begin{equation*}
    \begin{split}
       Cost Reduction = ((Total Requests - Executed Requests) \\ / Total Requests) * 100
    \end{split}
\end{equation*}

\noindent For RQ2, we measure the total rule hits (\textit{TotalHits}), applied or not applied rules (\textit{Applied} or \textit{NotApplied}), and their percentage rule coverages. A rule hit refers to a rule execution, which can be either applied or not applied. An applied rule refers to a fully executed (at least one time) rule. A not-applied rule relates to the partial execution of a rule which means a particular input (e.g., diagnose date), related to cancer messages, cannot be validated. A typical rule message, in this case, would be \textit{"This rule is not used because of diagnose date"}. This is a common case with validation rules (Figure~\ref{fig:sucessful-req}). 
The applied and not applied rule coverages are calculated as:
\begin{equation*}
    \begin{split}
       Coverage (Applied) = (Applied / Total Hits) * 100
    \end{split}
\end{equation*}
\begin{equation*}
    \begin{split}
       Coverage (NotApplied) = (NotApplied / Total Hits) * 100
    \end{split}
\end{equation*}
The rule coverage calculations reflect our perspective on the importance of rule execution. Specifically, we intend to emphasize the effectiveness of the generated test cases, where both applied and not applied rules are considered. The coverage metrics capture the proportion of applied (or not applied) rules relative to all rule hits (i.e., executed rules), but they do not account for never executed rules.

To reduce the effect of the randomness of the ML model of our approach and EvoMaster on the results, we repeated each experiment 30 times and performed statistical testing to check the significance of each difference between our approach and the default EvoMaster. We used the Mann-Whitney test as recommended in ~\cite{arcuri2011practical}. In addition, we relied on Vargha-Delaney's $\hat{A}_{12}$ to estimate the effect size (values of which range from 0 to 1) of the difference between the two approaches. A higher $\hat{A}_{12}$ value ($>0.5$) indicates our approach has a higher chance of yielding better results than the default EvoMaster, and vice versa. 


\subsection{Results and Discussion}
In the following section, we present the results and analyses of our evaluation corresponding to each RQ.

\subsubsection{RQ0 Results}\label{RQ0}
\begin{figure}[htbp]
  \centering
    \includegraphics[width=0.5\textwidth]{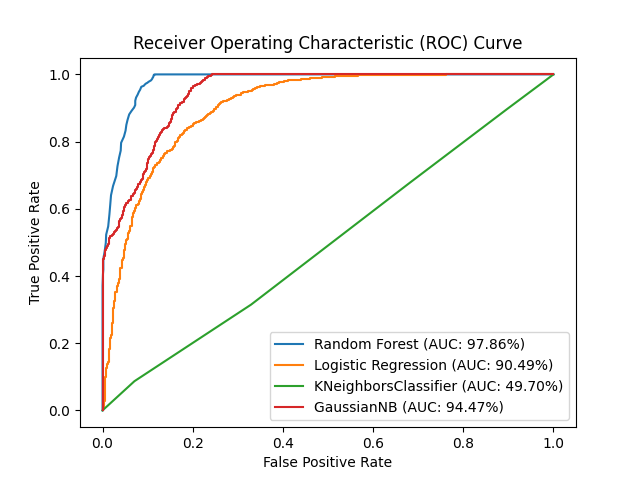}
  \caption{Performance comparison of the classifiers in ROC and AUC scores}
  \label{fig:performance_plot}
\end{figure}

Figure~\ref{fig:performance_plot} shows the performance results of the four classifiers as ROC and AUC. As shown in the figure, Random Forest performs better than Logistic Regression, K-Nearest Neighbors (KNN), and Gaussian Naive Bayes, as Random Forest achieves the highest AUC value, i.e., 97.86\%. Table~\ref{tab:comparison_report} summarizes each classifier's performance in terms of accuracy, precision, recall, and F1-score. Table~\ref{tab:comparison_report} shows that Random Forest achieved the highest accuracy, i.e., 95.40\%, which is at least 10\% higher than all the other three classifiers. Similarly, Random Forest outperforms the other three classifiers regarding precision, recall, and F1-score.

Notably, KNN demonstrates the weakest performance, which is due to the curse of dimensionality affecting its performance in high-dimensional spaces~\cite{pestov2013k}.
Based on these results, we opted for Random Forest and integrated it into {\method} for conducting other experiments for answering RQ1 and RQ2. 

\begin{table}[htbp]
\caption{Performance comparison of the classifiers in accuracy, precision, recall, and F1-score}
\centering
\begin{tabular}{lrrrr}
\toprule
    \textbf{Classifier} & 
    \textbf{Accuracy} &
    \textbf{Precision} & \textbf{Recall} & \textbf{F1-Score} \\
    \midrule
    Random Forest & 95.40\% & 96.38\% & 94.20\% & 95.05\% \\
    \rowcolor[HTML]{f9f9f9}
    Logistic Regression & 84.20\% & 85.39\% & 81.71\% & 82.77\% \\
    KNeighborsClassifier & 53.40\% & 49.78\% & 49.80\% & 48.46\% \\
    \rowcolor[HTML]{f9f9f9}
    GaussianNB & 75.20\% & 78.24\% & 78.32\% & 75.22\% \\
    \bottomrule
\end{tabular}
\label{tab:comparison_report}
\end{table}

\begin{Summary}
    \textbf{RQ0 Summary:} 
    Random Forest demonstrated superior performance across all metrics as compared to the other three classifiers. Notably, it achieved the highest AUC (97.86\%) in the ROC analysis.
\end{Summary}

\subsubsection{RQ1 Results}\label{RQ1}
Table~\ref{tab:rq1results} summarizes the results for each rule version under each environment. These results include the total number of generated requests (column \textit{Total Req.}), the total number of actually executed requests that are also predicted by our ML model to be successfully executed (Column \textit{Pred. Success}), the number of non-executed requests that have been filtered out by approach (column \textit{Pred. Failure}), and the number of executed requests that resulted in failures but predicted being successful (i.e., false positives, see column \textit{Pred. Success (F)}). 
In addition, we report results of accuracy, precision, recall, F1-score, and cost reduction. 

It can be observed that the overall cost reduction for all three environments and for all 10 versions is $\approx$31\%. This result implies that {\method} performs consistently well across the rule versions and across the different environments. In addition, a 31\% cost reduction is significant. For instance, for v1, {\method} managed to save cost by avoiding the execution of 14807 requests. The accuracy of {\method} ($\approx$91\%) is stable across the environments and versions, implying that the performance of {\method} does not degrade with the evolving versions in each environment. The overall precision and F1-score values are close to 87\% and 93\%, respectively. The recall values are 100\% in all cases, telling us that {\method} did not produce any false negatives, which is important to our context as a false negative implies not executing a request which might lead to the successful execution of a request.
The result also indicates that {\method}'s performance is stable despite rule changes across the rule versions. This is because their fundamental characteristics (e.g., data models and variables) remain relatively the same throughout the evolution of the rule set, often with only minor modifications. 


\begin{Summary}
    \textbf{RQ1 Summary:} {\method} has shown consistent and significant cost reduction across different GURI versions and environments. The average cost reduction achieved is approximately 31\%. 
\end{Summary}

\subsubsection{RQ2 Results}\label{RQ2}
Table~\ref{tab:rq2results} presents the results of the comparison between {\method} and the default EvoMaster (EM), in terms of the number of generated and executed requests, the total number of rule hits, the number of applied/not applied rules, and the coverage of applied/not applied rules. First, when looking at the total number of generated and executed requests, as already reported in RQ1, with {\method}, fewer requests were executed for each version under each environment when compared with the default EvoMaster; consequently, the overall cost is reduced. In terms of rule hits (i.e., the number of rules invoked in each request), which are positively correlated with the number of executed requests (with Pearson's correlation coefficients being 0.15 for the default EvoMaster and 0.30 for {\method}, respectively), obviously, the default EvoMaster achieved high numbers of rule hits for all the versions and under all the environments as it generated and executed more requests. Considering that the number of rules increases from v1 to v10 (Table \ref{tab:rules}), as expected, the number of rule hits also increases. For instance, rule hits increase from 533998 (v1) to 742003 (v10) under the development environment. 

Based on the execution results of the requests, we count the number of rules that are applied at least once and also fully executed (i.e., \textit{Rule Applied}), the number of validation rules that are partially executed as the condition part of such a rule is checked to be false (i.e., \textit{Rule NotApplied}). Notice that aggregation rules do not involve the \textit{Rule NotApplied} case. As shown in Table \ref{tab:rq2results}, due to more requests leading to more rule hits, the default EvoMaster achieved higher numbers of Rule Applied and Rule NotApplied instances for all the versions and across all the environments. However, when looking at the coverages of the Rule Applied and Rule NotApplied instances, {\method} performed very similarly to the default EvoMaster. This shows that our approach with a reduced number of requests can achieve the same level of coverage as EvoMaster. Interestingly, one can also observe that when evolving from v1 to v10, the coverage of Rule Applied decreases, and the coverage of Rule NotApplied increases. This is because the number and complexity of rules, especially validation rules, increased during the evolution from v1 to v10, as shown in Table \ref{tab:rules}. 

We also performed the Wilcoxon signed-rank test to check whether there exists a statistically significant difference between the default EvoMaster and {\method} in terms of the rule coverage. Results showed that the p-values are greater than 0.05, and $\hat{A}_{12}$ are around 0.5 for all cases. This indicates that there is insufficient evidence to conclude that there is a significant difference between the two approaches in terms of rule coverage. 

\begin{Summary}
    \textbf{RQ2 Summary:}
In terms of rule hits, the default EvoMaster outperformed {\method} due to generating and executing a higher number of requests. However, results show that {\method} can achieve the same rule coverage as the default EvoMaster with fewer executions. 
\end{Summary}




\subsection{Threats to Validity}\label{threats to validity}
Following, we discuss threats to the validity commonly reported in software engineering experiments~\cite{wohlin2012experimentation}.
To reduce threats to the \textit{external validity}, we used a real-world software application with ten versions that naturally evolved over four years of the operation of GURI, and deployed them under three environments. However, similar to many empirical software engineering studies, our results may not be generalizable to other application contexts, a common threat to the external validity~\cite{siegmund2015views}. 
To minimize threats to the \textit{internal validity}, we set up our experiment by following standard practices and recommended guidelines. Initially, we performed a pilot experiment to select a suitable ML classifier. 
We used a popular framework Optuna for hyperparameter tuning~\cite{optuna_2019}. 
We used the default/recommended parameters settings of EvoMaster~\cite{arcuri2019restful}. 
For the experiment setting, we set the number of repetitions to 30 and a one-hour fixed time budget for each run~\cite{arcuri2011practical,kim2022automated}. 
To handle threats to the \textit{construct validity}, we repeated our experiment 30 times to lower the effect of randomness. We analyzed experiment results using commonly used metrics (e.g., accuracy, precision). We compared {\method} with the default EvoMaster with the same set of metrics. In addition, we used the Mann-Whitney test and Vargha-Delaney's $\hat{A}_{12}$ effect size when comparing the two approaches, by following recommended guidelines~\cite{arcuri2011practical}, which reduces threats to the \textit{conclusion validity}. 

\begin{table*}[htbp]
 \ra{0.8}
\centering
\caption{Results of cost reduction of our approach and its' classifier's performance across the versions and environments}
\setlength{\extrarowheight}{0pt}
\addtolength{\extrarowheight}{\aboverulesep}
\addtolength{\extrarowheight}{\belowrulesep}
\setlength{\aboverulesep}{0pt}
\setlength{\belowrulesep}{0pt}
\scalebox{0.9}{
\begin{tabular}{llrrrrrrrrr} 
\toprule
\multicolumn{1}{c}{\textbf{Environment}} & \multicolumn{1}{c}{\textbf{V(i)}}       & \multicolumn{1}{c}{\textbf{Total Req.}}    & \multicolumn{1}{c}{\textbf{Pred. Succes}}  & \multicolumn{1}{c}{\textbf{Pred. Failure}} & \multicolumn{1}{c}{\textbf{Pred. Succ (Failure)}} & \multicolumn{1}{c}{\textbf{Accuracy}}       & \multicolumn{1}{c}{\textbf{Precision}}      & \multicolumn{1}{c}{\textbf{Recall}}       & \multicolumn{1}{c}{\textbf{F1-Score}}       & \multicolumn{1}{c}{\textbf{Cost Reduction (\%)}}  \\ 
\midrule
\multirow{10}{*}{\textit{Dev}}      & v1                                      & 47,471                                      & 32,664                                      & 14,807                                      & 4,137                                              & 91.28\%                                     & 87.33\%                                     & 100\%                                     & 93.23\%                                     & 31.19\%                                            \\
                                          & {\cellcolor[rgb]{0.973,0.973,0.973}}v2  & {\cellcolor[rgb]{0.973,0.973,0.973}}48,323 & {\cellcolor[rgb]{0.973,0.973,0.973}}33,432 & {\cellcolor[rgb]{0.973,0.973,0.973}}14,891 & {\cellcolor[rgb]{0.973,0.973,0.973}}4,295         & {\cellcolor[rgb]{0.973,0.973,0.973}}91.11\% & {\cellcolor[rgb]{0.973,0.973,0.973}}87.15\% & {\cellcolor[rgb]{0.973,0.973,0.973}}100\% & {\cellcolor[rgb]{0.973,0.973,0.973}}93.13\% & {\cellcolor[rgb]{0.973,0.973,0.973}}30.82\%        \\
                                          & v3                                      & 47,508                                     & 32,753                                     & 14,755                                     & 4,395                                             & 90.75\%                                     & 86.58\%                                     & 100\%                                     & 92.80\%                                     & 31.06\%                                            \\
                                          & {\cellcolor[rgb]{0.973,0.973,0.973}}v4  & {\cellcolor[rgb]{0.973,0.973,0.973}}47,020 & {\cellcolor[rgb]{0.973,0.973,0.973}}32,533 & {\cellcolor[rgb]{0.973,0.973,0.973}}14,487 & {\cellcolor[rgb]{0.973,0.973,0.973}}43 14         & {\cellcolor[rgb]{0.973,0.973,0.973}}90.83\% & {\cellcolor[rgb]{0.973,0.973,0.973}}86.73\% & {\cellcolor[rgb]{0.973,0.973,0.973}}100\% & {\cellcolor[rgb]{0.973,0.973,0.973}}92.90\% & {\cellcolor[rgb]{0.973,0.973,0.973}}30.81\%        \\
                                          & v5                                      & 37,078                                     & 25,577                                     & 11,501                                     & 3,307                                             & 91.08\%                                     & 87.07\%                                     & 100\%                                     & 93.08\%                                     & 31.02\%                                            \\
                                          & {\cellcolor[rgb]{0.973,0.973,0.973}}v6  & {\cellcolor[rgb]{0.973,0.973,0.973}}39,953 & {\cellcolor[rgb]{0.973,0.973,0.973}}27,378 & {\cellcolor[rgb]{0.973,0.973,0.973}}12,575 & {\cellcolor[rgb]{0.973,0.973,0.973}}3,614         & {\cellcolor[rgb]{0.973,0.973,0.973}}90.95\% & {\cellcolor[rgb]{0.973,0.973,0.973}}86.79\% & {\cellcolor[rgb]{0.973,0.973,0.973}}100\% & {\cellcolor[rgb]{0.973,0.973,0.973}}92.93\% & {\cellcolor[rgb]{0.973,0.973,0.973}}31.47\%        \\
                                          & v7                                      & 32,031                                     & 21,983                                     & 10,048                                     & 2,917                                             & 90.89\%                                     & 86.73\%                                     & 100\%                                     & 92.90\%                                     & 31.37\%                                            \\
                                          & {\cellcolor[rgb]{0.973,0.973,0.973}}v8  & {\cellcolor[rgb]{0.973,0.973,0.973}}32,108 & {\cellcolor[rgb]{0.973,0.973,0.973}}22,255 & {\cellcolor[rgb]{0.973,0.973,0.973}}9,853  & {\cellcolor[rgb]{0.973,0.973,0.973}}2,806         & {\cellcolor[rgb]{0.973,0.973,0.973}}91.26\% & {\cellcolor[rgb]{0.973,0.973,0.973}}87.39\% & {\cellcolor[rgb]{0.973,0.973,0.973}}100\% & {\cellcolor[rgb]{0.973,0.973,0.973}}93.27\% & {\cellcolor[rgb]{0.973,0.973,0.973}}30.69\%        \\
                                          & v9                                      & 33,866                                     & 23,303                                     & 10,563                                     & 3,078                                             & 90.91\%                                     & 86.79\%                                     & 100\%                                     & 92.92\%                                     & 31.19\%                                            \\
                                          & {\cellcolor[rgb]{0.973,0.973,0.973}}v10 & {\cellcolor[rgb]{0.973,0.973,0.973}}35,138 & {\cellcolor[rgb]{0.973,0.973,0.973}}24,217 & {\cellcolor[rgb]{0.973,0.973,0.973}}10,921 & {\cellcolor[rgb]{0.973,0.973,0.973}}3,105         & {\cellcolor[rgb]{0.973,0.973,0.973}}91.16\% & {\cellcolor[rgb]{0.973,0.973,0.973}}87.17\% & {\cellcolor[rgb]{0.973,0.973,0.973}}100\% & {\cellcolor[rgb]{0.973,0.973,0.973}}93.15\% & {\cellcolor[rgb]{0.973,0.973,0.973}}31.08\%       \\ 
\midrule
\multirow{10}{*}{\textit{Test}}           & v1                                      & 48,290                                     & 33,173                                     & 15,117                                     & 4,338                                             & 91.02\%                                     & 86.92\%                                     & 100\%                                     & 93.02\%                                     & 31.30\%                                            \\
                                          & {\cellcolor[rgb]{0.973,0.973,0.973}}v2  & {\cellcolor[rgb]{0.973,0.973,0.973}}47,854 & {\cellcolor[rgb]{0.973,0.973,0.973}}32,932 & {\cellcolor[rgb]{0.973,0.973,0.973}}14,922 & {\cellcolor[rgb]{0.973,0.973,0.973}}4,255         & {\cellcolor[rgb]{0.973,0.973,0.973}}91.11\% & {\cellcolor[rgb]{0.973,0.973,0.973}}87.11\% & {\cellcolor[rgb]{0.973,0.973,0.973}}100\% & {\cellcolor[rgb]{0.973,0.973,0.973}}92.96\% & {\cellcolor[rgb]{0.973,0.973,0.973}}31.18\%        \\
                                          & v3                                      & 44,010                                     & 30,239                                     & 13,771                                     & 3,945                                             & 91.03\%                                     & 86.96\%                                     & 100\%                                     & 93.38\%                                     & 31.29\%                                            \\
                                          & {\cellcolor[rgb]{0.973,0.973,0.973}}v4  & {\cellcolor[rgb]{0.973,0.973,0.973}}41,231 & {\cellcolor[rgb]{0.973,0.973,0.973}}28,382 & {\cellcolor[rgb]{0.973,0.973,0.973}}12,849 & {\cellcolor[rgb]{0.973,0.973,0.973}}3,635         & {\cellcolor[rgb]{0.973,0.973,0.973}}91.18\% & {\cellcolor[rgb]{0.973,0.973,0.973}}87.19\% & {\cellcolor[rgb]{0.973,0.973,0.973}}100\% & {\cellcolor[rgb]{0.973,0.973,0.973}}92.29\% & {\cellcolor[rgb]{0.973,0.973,0.973}}31.17\%        \\
                                          & v5                                      & 55,488                                     & 38,164                                     & 17,324                                     & 4,940                                             & 91.09\%                                     & 87.19\%                                     & 100\%                                     & 92.29\%                                     & 31.22\%                                            \\
                                          & {\cellcolor[rgb]{0.973,0.973,0.973}}v6  & {\cellcolor[rgb]{0.973,0.973,0.973}}45,569 & {\cellcolor[rgb]{0.973,0.973,0.973}}31,551 & {\cellcolor[rgb]{0.973,0.973,0.973}}14,018 & {\cellcolor[rgb]{0.973,0.973,0.973}}4,080         & {\cellcolor[rgb]{0.973,0.973,0.973}}91.04\% & {\cellcolor[rgb]{0.973,0.973,0.973}}87.01\% & {\cellcolor[rgb]{0.973,0.973,0.973}}100\% & {\cellcolor[rgb]{0.973,0.973,0.973}}93.22\% & {\cellcolor[rgb]{0.973,0.973,0.973}}30.76\%        \\
                                          & v7                                      & 55,287                                     & 37,914                                     & 17,373                                     & 4,936                                             & 91.07\%                                     & 87.12\%                                     & 100\%                                     & 92.40\%                                     & 31.42\%                                            \\
                                          & {\cellcolor[rgb]{0.973,0.973,0.973}}v8  & {\cellcolor[rgb]{0.973,0.973,0.973}}54,149 & {\cellcolor[rgb]{0.973,0.973,0.973}}37,302 & {\cellcolor[rgb]{0.973,0.973,0.973}}16,847 & {\cellcolor[rgb]{0.973,0.973,0.973}}4,896         & {\cellcolor[rgb]{0.973,0.973,0.973}}91.78\% & {\cellcolor[rgb]{0.973,0.973,0.973}}86.51\% & {\cellcolor[rgb]{0.973,0.973,0.973}}100\% & {\cellcolor[rgb]{0.973,0.973,0.973}}93.07\% & {\cellcolor[rgb]{0.973,0.973,0.973}}31.11\%        \\
                                          & v9                                      & 43,884                                     & 30,228                                     & 13,655                                     & 4,004                                             & 91.87\%                                     & 86.78\%                                     & 100\%                                     & 93.37\%                                     & 31.12\%                                            \\
                                          & {\cellcolor[rgb]{0.973,0.973,0.973}}v10 & {\cellcolor[rgb]{0.973,0.973,0.973}}53,261 & {\cellcolor[rgb]{0.973,0.973,0.973}}36,693 & {\cellcolor[rgb]{0.973,0.973,0.973}}16,568 & {\cellcolor[rgb]{0.973,0.973,0.973}}4,801         & {\cellcolor[rgb]{0.973,0.973,0.973}}90.98\% & {\cellcolor[rgb]{0.973,0.973,0.973}}86.92\% & {\cellcolor[rgb]{0.973,0.973,0.973}}100\% & {\cellcolor[rgb]{0.973,0.973,0.973}}93.35\% & {\cellcolor[rgb]{0.973,0.973,0.973}}31.10\%      \\ 
\midrule
\multirow{10}{*}{\textit{Prod}}           & v1                                      & 41,211                                     & 28,411                                     & 12,800                                     & 3,730                                             & 90.95\%                                     & 86.87\%                                     & 100\%                                     & 92.97\%                                     & 31.05\%                                            \\
                                          & {\cellcolor[rgb]{0.973,0.973,0.973}}v2  & {\cellcolor[rgb]{0.973,0.973,0.973}}33,584 & {\cellcolor[rgb]{0.973,0.973,0.973}}23,169 & {\cellcolor[rgb]{0.973,0.973,0.973}}10,415 & {\cellcolor[rgb]{0.973,0.973,0.973}}3,021         & {\cellcolor[rgb]{0.973,0.973,0.973}}91.00\%    & {\cellcolor[rgb]{0.973,0.973,0.973}}86.96\% & {\cellcolor[rgb]{0.973,0.973,0.973}}100\% & {\cellcolor[rgb]{0.973,0.973,0.973}}93.02\% & {\cellcolor[rgb]{0.973,0.973,0.973}}31.00\%           \\
                                          & v3                                      & 43,137                                     & 29,870                                     & 13,267                                     & 3,797                                             & 91.20\%                                     & 87.29\%                                     & 100\%                                     & 93.21\%                                     & 30.75\%                                            \\
                                          & {\cellcolor[rgb]{0.973,0.973,0.973}}v4  & {\cellcolor[rgb]{0.973,0.973,0.973}}33,370 & {\cellcolor[rgb]{0.973,0.973,0.973}}22,930 & {\cellcolor[rgb]{0.973,0.973,0.973}}10,440 & {\cellcolor[rgb]{0.973,0.973,0.973}}2,989         & {\cellcolor[rgb]{0.973,0.973,0.973}}91.05\% & {\cellcolor[rgb]{0.973,0.973,0.973}}86.96\% & {\cellcolor[rgb]{0.973,0.973,0.973}}100\% & {\cellcolor[rgb]{0.973,0.973,0.973}}93.02\% & {\cellcolor[rgb]{0.973,0.973,0.973}}31.28\%        \\
                                          & v5                                      & 38,077                                     & 26,192                                     & 11,885                                     & 3,444                                             & 90.95\%                                     & 86.85\%                                     & 100\%                                     & 92.96\%                                     & 31.21\%                                            \\
                                          & {\cellcolor[rgb]{0.973,0.973,0.973}}v6  & {\cellcolor[rgb]{0.973,0.973,0.973}}37,938 & {\cellcolor[rgb]{0.973,0.973,0.973}}26,184 & {\cellcolor[rgb]{0.973,0.973,0.973}}11,754 & {\cellcolor[rgb]{0.973,0.973,0.973}}3,469         & {\cellcolor[rgb]{0.973,0.973,0.973}}90.85\% & {\cellcolor[rgb]{0.973,0.973,0.973}}86.75\% & {\cellcolor[rgb]{0.973,0.973,0.973}}100\% & {\cellcolor[rgb]{0.973,0.973,0.973}}92.90\% & {\cellcolor[rgb]{0.973,0.973,0.973}}30.98\%        \\
                                          & v7                                      & 35,500                                     & 24,545                                     & 10,955                                     & 3,151                                             & 91.12\%                                     & 87.16\%                                     & 100\%                                     & 93.14\%                                     & 30.85\%                                            \\
                                          & {\cellcolor[rgb]{0.973,0.973,0.973}}v8  & {\cellcolor[rgb]{0.973,0.973,0.973}}32,412 & {\cellcolor[rgb]{0.973,0.973,0.973}}22,330 & {\cellcolor[rgb]{0.973,0.973,0.973}}10,082 & {\cellcolor[rgb]{0.973,0.973,0.973}}2,989         & {\cellcolor[rgb]{0.973,0.973,0.973}}90.77\% & {\cellcolor[rgb]{0.973,0.973,0.973}}86.61\% & {\cellcolor[rgb]{0.973,0.973,0.973}}100\% & {\cellcolor[rgb]{0.973,0.973,0.973}}92.82\% & {\cellcolor[rgb]{0.973,0.973,0.973}}31.11\%        \\
                                          & v9                                      & 36,148                                     & 24,946                                     & 11,202                                     & 3,255                                             & 91.00\%                                        & 86.95\%                                     & 100\%                                     & 93.02\%                                     & 30.98\%                                            \\
                                          & {\cellcolor[rgb]{0.973,0.973,0.973}}v10 & {\cellcolor[rgb]{0.973,0.973,0.973}}39,047 & {\cellcolor[rgb]{0.973,0.973,0.973}}26,913 & {\cellcolor[rgb]{0.973,0.973,0.973}}12,134 & {\cellcolor[rgb]{0.973,0.973,0.973}}3,462         & {\cellcolor[rgb]{0.973,0.973,0.973}}91.13\% & {\cellcolor[rgb]{0.973,0.973,0.973}}87.13\% & {\cellcolor[rgb]{0.973,0.973,0.973}}100\% & {\cellcolor[rgb]{0.973,0.973,0.973}}93.12\% & {\cellcolor[rgb]{0.973,0.973,0.973}}31.07\%         \\ 
\bottomrule
\multicolumn{1}{c}{}                      & \multicolumn{1}{l}{}                    &                                            &                                            &                                            &                                                   &                                             &                                             &                                           &                                             & \multicolumn{1}{r}{}                              
\end{tabular}
}
\label{tab:rq1results}
\end{table*}

\begin{table*}[htbp]
 \ra{0.8}
\caption{Approach Comparison based on Rule Hits and Coverage Across Versions and Environments}
\centering
\setlength{\extrarowheight}{0pt}
\addtolength{\extrarowheight}{\aboverulesep}
\addtolength{\extrarowheight}{\belowrulesep}
\setlength{\aboverulesep}{1pt}
\setlength{\belowrulesep}{1pt}
\scalebox{0.9}{
\begin{tabular}{llrrrrrrrrrrrr} 
\toprule
\textbf{Env.} & \multicolumn{1}{r}{\textbf{V(i)}} & \multicolumn{2}{c}{\textbf{Requests}}                        & \multicolumn{2}{c}{\textbf{Rule Hits}}                             & \multicolumn{2}{c}{\textbf{Rule Applied}}                            & \multicolumn{2}{c}{\textbf{Rule Not Applied}}                              & \multicolumn{2}{c}{\textbf{Coverage Applied}}                         & \multicolumn{2}{c}{\textbf{Coverage Not Applied}}  \\ 
\cmidrule(l{1pt}r{1pt}){3-4}
\cmidrule(l{1pt}r{1pt}){5-6}
\cmidrule(l{1pt}r{1pt}){7-8}
\cmidrule(l{1pt}r{1pt}){9-10}
\cmidrule(l{1pt}r{1pt}){11-12}
\cmidrule(l{1pt}r{1pt}){13-14}

                                & \multicolumn{1}{r}{}                      & {\cellcolor[rgb]{0.973,0.973,0.973}}Our App.                                   & EM     & {\cellcolor[rgb]{0.973,0.973,0.973}}Our App.                                    & EM        & {\cellcolor[rgb]{0.973,0.973,0.973}}Our App.                                   & EM      & {\cellcolor[rgb]{0.973,0.973,0.973}}Our App.                                    & EM        & {\cellcolor[rgb]{0.973,0.973,0.973}}Our App.                                    & EM      & {\cellcolor[rgb]{0.973,0.973,0.973}}Our App.                                    & EM                \\ 
\midrule
\multirow{10}{*}{\textit{Dev}} & v1                                                                       & {\cellcolor[rgb]{0.973,0.973,0.973}}32,664 & 93,558 & {\cellcolor[rgb]{0.973,0.973,0.973}}533,998 & 1,053,641 & {\cellcolor[rgb]{0.973,0.973,0.973}}77,259 & 158,134 & {\cellcolor[rgb]{0.973,0.973,0.973}}456,739 & 895,507   & {\cellcolor[rgb]{0.973,0.973,0.973}}14.47\% & 15.01\% & {\cellcolor[rgb]{0.973,0.973,0.973}}85.53\% & 84.99\%           \\
                                & v2                                                                       & {\cellcolor[rgb]{0.973,0.973,0.973}}33,432 & 78,773 & {\cellcolor[rgb]{0.973,0.973,0.973}}561,721 & 894,402   & {\cellcolor[rgb]{0.973,0.973,0.973}}82,266 & 118,778 & {\cellcolor[rgb]{0.973,0.973,0.973}}479,455 & 775,624   & {\cellcolor[rgb]{0.973,0.973,0.973}}14.65\% & 13.28\% & {\cellcolor[rgb]{0.973,0.973,0.973}}85.35\% & 86.72\%           \\
                                & v3                                                                       & {\cellcolor[rgb]{0.973,0.973,0.973}}32,753 & 63,159 & {\cellcolor[rgb]{0.973,0.973,0.973}}741,482 & 995,216   & {\cellcolor[rgb]{0.973,0.973,0.973}}85,075 & 110,074 & {\cellcolor[rgb]{0.973,0.973,0.973}}656,407 & 885,142   & {\cellcolor[rgb]{0.973,0.973,0.973}}11.47\% & 11.06\% & {\cellcolor[rgb]{0.973,0.973,0.973}}88.53\% & 88.94\%           \\
                                & v4                                                                       & {\cellcolor[rgb]{0.973,0.973,0.973}}32,533 & 72,354 & {\cellcolor[rgb]{0.973,0.973,0.973}}738,539 & 1,150,109 & {\cellcolor[rgb]{0.973,0.973,0.973}}82,132 & 129,862 & {\cellcolor[rgb]{0.973,0.973,0.973}}656,407 & 1,020,247 & {\cellcolor[rgb]{0.973,0.973,0.973}}11.12\% & 11.29\% & {\cellcolor[rgb]{0.973,0.973,0.973}}88.88\% & 88.71\%           \\
                                & v5                                                                       & {\cellcolor[rgb]{0.973,0.973,0.973}}25,577 & 44,874 & {\cellcolor[rgb]{0.973,0.973,0.973}}637,463 & 773,199   & {\cellcolor[rgb]{0.973,0.973,0.973}}64,888 & 81,775  & {\cellcolor[rgb]{0.973,0.973,0.973}}572,575 & 691,424   & {\cellcolor[rgb]{0.973,0.973,0.973}}10.18\% & 10.58\% & {\cellcolor[rgb]{0.973,0.973,0.973}}89.82\% & 89.42\%           \\
                                & v6                                                                       & {\cellcolor[rgb]{0.973,0.973,0.973}}27,378 & 79,899 & {\cellcolor[rgb]{0.973,0.973,0.973}}687,797 & 1,397,483 & {\cellcolor[rgb]{0.973,0.973,0.973}}69,933 & 141,945 & {\cellcolor[rgb]{0.973,0.973,0.973}}617,864 & 1,255,538 & {\cellcolor[rgb]{0.973,0.973,0.973}}10.17\% & 10.16\% & {\cellcolor[rgb]{0.973,0.973,0.973}}89.83\% & 89.84\%           \\
                                & v7                                                                       & {\cellcolor[rgb]{0.973,0.973,0.973}}21,983 & 84,855 & {\cellcolor[rgb]{0.973,0.973,0.973}}648,377 & 1,748,928 & {\cellcolor[rgb]{0.973,0.973,0.973}}55,051 & 154,161 & {\cellcolor[rgb]{0.973,0.973,0.973}}593,326 & 1,594,767 & {\cellcolor[rgb]{0.973,0.973,0.973}}8.49\%  & 8.81\%  & {\cellcolor[rgb]{0.973,0.973,0.973}}91.51\% & 91.19\%           \\
                                & v8                                                                       & {\cellcolor[rgb]{0.973,0.973,0.973}}22,255 & 49,498 & {\cellcolor[rgb]{0.973,0.973,0.973}}679,334 & 1,012,832 & {\cellcolor[rgb]{0.973,0.973,0.973}}55,650 & 86,739  & {\cellcolor[rgb]{0.973,0.973,0.973}}623,684 & 926,093   & {\cellcolor[rgb]{0.973,0.973,0.973}}8.19\%  & 8.56\%  & {\cellcolor[rgb]{0.973,0.973,0.973}}91.81\% & 91.44\%           \\
                                & v9                                                                       & {\cellcolor[rgb]{0.973,0.973,0.973}}23,303 & 51,187 & {\cellcolor[rgb]{0.973,0.973,0.973}}716,937 & 1,050,210 & {\cellcolor[rgb]{0.973,0.973,0.973}}58,092 & 89,670  & {\cellcolor[rgb]{0.973,0.973,0.973}}658,845 & 960,540   & {\cellcolor[rgb]{0.973,0.973,0.973}}8.10\%  & 8.54\%  & {\cellcolor[rgb]{0.973,0.973,0.973}}91.90\% & 91.46\%           \\
                                & v10                                                                      & {\cellcolor[rgb]{0.973,0.973,0.973}}24,217 & 43,823 & {\cellcolor[rgb]{0.973,0.973,0.973}}742,003 & 933,208   & {\cellcolor[rgb]{0.973,0.973,0.973}}61,241 & 73,042  & {\cellcolor[rgb]{0.973,0.973,0.973}}680,762 & 860,166   & {\cellcolor[rgb]{0.973,0.973,0.973}}8.25\%  & 7.83\%  & {\cellcolor[rgb]{0.973,0.973,0.973}}91.75\% & 92.17\%            \\ 
\midrule
\multirow{10}{*}{\textit{Test}} & 
                                 v1                                                                       & {\cellcolor[rgb]{0.973,0.973,0.973}}33,173 & 89,896 & {\cellcolor[rgb]{0.973,0.973,0.973}}336,318 & 628,806   & {\cellcolor[rgb]{0.973,0.973,0.973}}74,133 & 135,978 & {\cellcolor[rgb]{0.973,0.973,0.973}}262,185 & 492,828   & {\cellcolor[rgb]{0.973,0.973,0.973}}22.04\% & 21.62\% & {\cellcolor[rgb]{0.973,0.973,0.973}}77.96\% & 78.38\%           \\
                                & v2                                                                       & {\cellcolor[rgb]{0.973,0.973,0.973}}32,932 & 95,042 & {\cellcolor[rgb]{0.973,0.973,0.973}}348,270 & 694,370   & {\cellcolor[rgb]{0.973,0.973,0.973}}75,678 & 151,389 & {\cellcolor[rgb]{0.973,0.973,0.973}}272,589 & 544,981   & {\cellcolor[rgb]{0.973,0.973,0.973}}21.73\% & 21.81\% & {\cellcolor[rgb]{0.973,0.973,0.973}}78.27\% & 78.19\%           \\
                                & v3                                                                       & {\cellcolor[rgb]{0.973,0.973,0.973}}30,239 & 88,576 & {\cellcolor[rgb]{0.973,0.973,0.973}}472,119 & 928,909   & {\cellcolor[rgb]{0.973,0.973,0.973}}70,034 & 145,070 & {\cellcolor[rgb]{0.973,0.973,0.973}}402,085 & 783,839   & {\cellcolor[rgb]{0.973,0.973,0.973}}14.83\% & 15.97\% & {\cellcolor[rgb]{0.973,0.973,0.973}}85.17\% & 84.03\%           \\
                                & v4                                                                       & {\cellcolor[rgb]{0.973,0.973,0.973}}28,382 & 80,103 & {\cellcolor[rgb]{0.973,0.973,0.973}}438,736 & 841,575   & {\cellcolor[rgb]{0.973,0.973,0.973}}66,337 & 130,867 & {\cellcolor[rgb]{0.973,0.973,0.973}}372,399 & 710,708   & {\cellcolor[rgb]{0.973,0.973,0.973}}15.15\% & 15.55\% & {\cellcolor[rgb]{0.973,0.973,0.973}}84.85\% & 84.45\%           \\
                                & v5                                                                       & {\cellcolor[rgb]{0.973,0.973,0.973}}38,164 & 78,392 & {\cellcolor[rgb]{0.973,0.973,0.973}}639,976 & 889,586   & {\cellcolor[rgb]{0.973,0.973,0.973}}89,512 & 130,082 & {\cellcolor[rgb]{0.973,0.973,0.973}}550,464 & 759,504   & {\cellcolor[rgb]{0.973,0.973,0.973}}13.98\% & 14.62\% & {\cellcolor[rgb]{0.973,0.973,0.973}}86.02\% & 85.38\%           \\
                                & v6                                                                       & {\cellcolor[rgb]{0.973,0.973,0.973}}31,551 & 91,252 & {\cellcolor[rgb]{0.973,0.973,0.973}}550,913 & 1,088,872 & {\cellcolor[rgb]{0.973,0.973,0.973}}76,208 & 151,298 & {\cellcolor[rgb]{0.973,0.973,0.973}}474,705 & 937,574   & {\cellcolor[rgb]{0.973,0.973,0.973}}13.83\% & 13.89\% & {\cellcolor[rgb]{0.973,0.973,0.973}}86.17\% & 86.11\%           \\
                                & v7                                                                       & {\cellcolor[rgb]{0.973,0.973,0.973}}37,914 & 78,168 & {\cellcolor[rgb]{0.973,0.973,0.973}}812,660 & 1,135,834 & {\cellcolor[rgb]{0.973,0.973,0.973}}86,802 & 128,884 & {\cellcolor[rgb]{0.973,0.973,0.973}}725,858 & 1,006,950 & {\cellcolor[rgb]{0.973,0.973,0.973}}10.68\% & 11.34\% & {\cellcolor[rgb]{0.973,0.973,0.973}}89.32\% & 88.66\%           \\
                                & v8                                                                       & {\cellcolor[rgb]{0.973,0.973,0.973}}37,302 & 76,436 & {\cellcolor[rgb]{0.973,0.973,0.973}}805,472 & 1,150,039 & {\cellcolor[rgb]{0.973,0.973,0.973}}92,333 & 124,791 & {\cellcolor[rgb]{0.973,0.973,0.973}}713,139 & 1,025,248 & {\cellcolor[rgb]{0.973,0.973,0.973}}11.46\% & 10.85\% & {\cellcolor[rgb]{0.973,0.973,0.973}}88.54\% & 89.15\%           \\
                                & v9                                                                       & {\cellcolor[rgb]{0.973,0.973,0.973}}30,228 & 93,807 & {\cellcolor[rgb]{0.973,0.973,0.973}}662,471 & 1,411,984 & {\cellcolor[rgb]{0.973,0.973,0.973}}73,626 & 152,027 & {\cellcolor[rgb]{0.973,0.973,0.973}}588,845 & 1,259,957 & {\cellcolor[rgb]{0.973,0.973,0.973}}11.11\% & 10.76\% & {\cellcolor[rgb]{0.973,0.973,0.973}}88.89\% & 89.24\%           \\
                                & v10                                                                      & {\cellcolor[rgb]{0.973,0.973,0.973}}36,693 & 91,002 & {\cellcolor[rgb]{0.973,0.973,0.973}}814,143 & 1,378,728 & {\cellcolor[rgb]{0.973,0.973,0.973}}88,676 & 145,903 & {\cellcolor[rgb]{0.973,0.973,0.973}}725,467 & 1,232,825 & {\cellcolor[rgb]{0.973,0.973,0.973}}10.89\% & 10.58\% & {\cellcolor[rgb]{0.973,0.973,0.973}}89.11\% & 89.42\%           \\ 
\midrule
\multirow{10}{*}{\textit{Prod}} & v1                                                                       & {\cellcolor[rgb]{0.973,0.973,0.973}}28,411 & 93,972 & {\cellcolor[rgb]{0.973,0.973,0.973}}285,966 & 661,030   & {\cellcolor[rgb]{0.973,0.973,0.973}}64,118 & 145,123 & {\cellcolor[rgb]{0.973,0.973,0.973}}221,848 & 515,907   & {\cellcolor[rgb]{0.973,0.973,0.973}}22.42\% & 21.95\% & {\cellcolor[rgb]{0.973,0.973,0.973}}77.58\% & 78.05\%           \\
                                & v2                                                                       & {\cellcolor[rgb]{0.973,0.973,0.973}}23,169 & 85,055 & {\cellcolor[rgb]{0.973,0.973,0.973}}250,632 & 623,840   & {\cellcolor[rgb]{0.973,0.973,0.973}}53,848 & 134,858 & {\cellcolor[rgb]{0.973,0.973,0.973}}196,784 & 488,982   & {\cellcolor[rgb]{0.973,0.973,0.973}}21.48\% & 21.62\% & {\cellcolor[rgb]{0.973,0.973,0.973}}78.52\% & 78.38\%           \\
                                & v3                                                                       & {\cellcolor[rgb]{0.973,0.973,0.973}}29,870 & 86,166 & {\cellcolor[rgb]{0.973,0.973,0.973}}443,170 & 902,315   & {\cellcolor[rgb]{0.973,0.973,0.973}}71,304 & 138,218 & {\cellcolor[rgb]{0.973,0.973,0.973}}371,866 & 764,097   & {\cellcolor[rgb]{0.973,0.973,0.973}}16.09\% & 15.32\% & {\cellcolor[rgb]{0.973,0.973,0.973}}83.91\% & 84.68\%           \\
                                & v4                                                                       & {\cellcolor[rgb]{0.973,0.973,0.973}}22,930 & 96,528 & {\cellcolor[rgb]{0.973,0.973,0.973}}340,201 & 1,019,712 & {\cellcolor[rgb]{0.973,0.973,0.973}}51,914 & 154,461 & {\cellcolor[rgb]{0.973,0.973,0.973}}288,287 & 865,251   & {\cellcolor[rgb]{0.973,0.973,0.973}}15.26\% & 15.15\% & {\cellcolor[rgb]{0.973,0.973,0.973}}84.74\% & 84.85\%           \\
                                & v5                                                                       & {\cellcolor[rgb]{0.973,0.973,0.973}}26,192 & 81,454 & {\cellcolor[rgb]{0.973,0.973,0.973}}437,591 & 930,347   & {\cellcolor[rgb]{0.973,0.973,0.973}}61,894 & 131,708 & {\cellcolor[rgb]{0.973,0.973,0.973}}375,697 & 798,639   & {\cellcolor[rgb]{0.973,0.973,0.973}}14.14\% & 14.16\% & {\cellcolor[rgb]{0.973,0.973,0.973}}85.86\% & 85.84\%           \\
                                & v6                                                                       & {\cellcolor[rgb]{0.973,0.973,0.973}}26,184 & 92,862 & {\cellcolor[rgb]{0.973,0.973,0.973}}468,305 & 1,126,310 & {\cellcolor[rgb]{0.973,0.973,0.973}}62,771 & 153,969 & {\cellcolor[rgb]{0.973,0.973,0.973}}405,534 & 972,341   & {\cellcolor[rgb]{0.973,0.973,0.973}}13.40\% & 13.67\% & {\cellcolor[rgb]{0.973,0.973,0.973}}86.60\% & 86.33\%           \\
                                & v7                                                                       & {\cellcolor[rgb]{0.973,0.973,0.973}}24,545 & 94,657 & {\cellcolor[rgb]{0.973,0.973,0.973}}527,147 & 1,363,694 & {\cellcolor[rgb]{0.973,0.973,0.973}}60,764 & 154,658 & {\cellcolor[rgb]{0.973,0.973,0.973}}466,383 & 1,209,036 & {\cellcolor[rgb]{0.973,0.973,0.973}}11.53\% & 11.34\% & {\cellcolor[rgb]{0.973,0.973,0.973}}88.47\% & 88.66\%           \\
                                & v8                                                                       & {\cellcolor[rgb]{0.973,0.973,0.973}}22,330 & 66,891 & {\cellcolor[rgb]{0.973,0.973,0.973}}508,787 & 985,497   & {\cellcolor[rgb]{0.973,0.973,0.973}}53,359 & 110,392 & {\cellcolor[rgb]{0.973,0.973,0.973}}455,428 & 875,105   & {\cellcolor[rgb]{0.973,0.973,0.973}}10.49\% & 11.20\% & {\cellcolor[rgb]{0.973,0.973,0.973}}89.51\% & 88.80\%           \\
                                & v9                                                                       & {\cellcolor[rgb]{0.973,0.973,0.973}}24,946 & 90,913 & {\cellcolor[rgb]{0.973,0.973,0.973}}551,355 & 1,372,616 & {\cellcolor[rgb]{0.973,0.973,0.973}}59,557 & 149,276 & {\cellcolor[rgb]{0.973,0.973,0.973}}491,798 & 1,223,340 & {\cellcolor[rgb]{0.973,0.973,0.973}}10.80\% & 10.88\% & {\cellcolor[rgb]{0.973,0.973,0.973}}89.20\% & 89.12\%           \\
                                & v10                                                                      & {\cellcolor[rgb]{0.973,0.973,0.973}}26,913 & 69,905 & {\cellcolor[rgb]{0.973,0.973,0.973}}608,506 & 1,060,819 & {\cellcolor[rgb]{0.973,0.973,0.973}}62,900 & 113,411 & {\cellcolor[rgb]{0.973,0.973,0.973}}545,606 & 947,408   & {\cellcolor[rgb]{0.973,0.973,0.973}}10.34\% & 10.69\% & {\cellcolor[rgb]{0.973,0.973,0.973}}89.66\% & 89.31\%          \\
\bottomrule
\end{tabular}}
\label{tab:rq2results}
\end{table*}

\section{Experiences and Lessons Learned}\label{sec:lessons}

\textbf{Generalizability:} Even though we extended EvoMaster with an ML classifier, such a classifier can be integrated into other REST API-based tools (e.g., RESTest~\cite{martin2020restest}, RESTler~\cite{atlidakis2019restler}, and RestTestGen~\cite{corradini2022automated}). As a result, testing cost reduction can be achieved together with testing strategies implemented by these tools, such as Adaptive Random Testing and Constraint-based Testing. Moreover, in our context, Random Forest showed the best results. Other classifiers may be better in other contexts, which can be integrated into EvoMaster and other classifiers in the future. 

For now, we experimented with one sub-system of CaReSS, i.e., GURI. Our experiment results show that we save around 30\% of the testing cost by simply introducing an ML classifier. This result is very encouraging. Therefore, as the next step, we will perform additional experiments with other sub-systems of CaReSS and also CaReSS as a whole. Naturally, the implementation of {\method} can be reused for extended experiments. Based on the encouraging results of the current experiments, we expect that at least a similar testing effort can be saved. Furthermore, a large-scale empirical study will be needed to see whether Random Forrest performs best when testing other sub-systems. Additionally, this empirical study could provide valuable insights into the impact of various factors such as dataset size, preprocessing steps, and hyperparameter tuning that would most likely change alongside the selected classifier.

{\method} also holds the potential for broader applicability, extending beyond its current application domain; while our experiments were centered on GURI within the CaReSS system, the method's core principles can be utilized in testing REST APIs in other domains, e.g., healthcare IoT~\cite{sartaj2023testing}. The applicability of our method requires configuring application-specific details, such as OAS schema and data pre-processing. 

\textbf{Test case and dataset quality:} Many studies emphasize the importance of collecting high-quality and diverse datasets to train machine learning models effectively~\cite{jain2020overview, cortes1994limits, lu2023machine}. A comprehensive dataset in terms of API requests and responses from various scenarios (i.e., test cases) ensures better performance of the ML model. However, this becomes a challenge when we look at existing testing tools' limitations (e.g., generation of pseudo-random inputs and inter-parameter dependencies)~\cite{martin2021black, lin2022forest}. These tools lack the capability to generate domain-specific data, e.g., medical data, which limits the availability of diverse and representative test cases. In our work, we address this challenge by leveraging synthetic data, which, though not reflecting real-world scenarios, provides an alternative for training ML models when domain-specific data is unavailable. In our case, it was impossible to get real data due to the legislation of the General Data Protection Regulation (GDPR) from the European Union; therefore, we had to generate data for ML training ourselves. As a result, how to train ML models when real datasets contain personal data and are restricted by GDPR is an interesting area of research.

\textbf{Balancing testing cost and effectiveness:} ML-based approaches can provide significant savings in testing costs, but it is as much of important to strike a balance between cost reduction and maintaining the effectiveness of the testing process. In our case, the effectiveness is measured by the rule coverage, and the results show that we are maintaining similar effectiveness as default EvoMaster while reducing costs. In this regard, finding and optimizing the trade-off between false positives and false negatives are valuable lessons learned in this study. 
False positives occur when the model incorrectly predicts a successful API request as positive, while false negatives happen when the model wrongly predicts an unsuccessful request as negative. These occurrences affect precision and recall scores, as shown in Table~\ref{tab:rq1results}.

\textbf{Maintainability:} There are several dimensions of maintainability. First, {\method} can be adapted and integrated into other existing REST API-based testing tools. However, it is important to acknowledge that these tools are prone to continuous updates and changes which would require careful consideration in terms of compatibility and usability. Second, for the subject application, specifically, REST APIs under test, if changes in the respective OAS schema are of major importance, they need to be addressed accordingly in the approach, following each phase as shown in Figure~\ref{fig:approach}. Finally, after a certain period of time, the trained ML model would need an update. To this end, approaches such as transfer learning can be considered to update the ML model~\cite{chengjie2023evoclinical}. 

\section{Related Work}\label{relatedworks}

With the widespread deployment and usage of web APIs based applications, testing them to ensure their quality thoroughly is crucial. As a result, numerous automated techniques and tools have emerged in recent years for testing REST APIs such as EvoMaster~\cite{arcuri2018evomaster, arcuri2019restful, arcuri2023building}, RESTest~\cite{martin2020restest}, RESTler~\cite{atlidakis2019restler}, RestTestGen~\cite{corradini2022automated, corradini2022resttestgen}, bBOXRT~\cite{laranjeiro2021black}, Schemathesis~\cite{hatfield2022deriving}, Dredd\footnote{https://github.com/apiaryio/dredd/}, Tcases\footnote{https://github.com/Cornutum/tcases/tree/master/tcases-openapi/}, and APIFuzzer\footnote{https://github.com/KissPeter/APIFuzzer/}.
Several empirical studies have also been performed, providing deeper insights (e.g., coverage, performance, and fault detection) into the strengths and limitations of the aforementioned automated testing tools~\cite{corradini2021empirical, hatfield2022deriving, kim2022automated}. Despite of these insights, {\method}, even though presented as an extension of EvoMaster, can seamlessly be integrated with any other REST API-based tool.

Automated testing techniques for RESTful APIs primarily rely on black-box testing, generating test inputs randomly from the API specification. However, these approaches often generate considerable "invalid" inputs, leading to unsuccessful HTTP calls. Consequently, these randomly generated inputs fail to simulate real-world scenarios accurately. Additionally, the existing literature overlooks the testing of REST API functionalities beyond input validity, neglecting the evaluation of meaningful responses, i.e., successful requests~\cite{ed2018automatic, golmohammadi2022testing, viglianisi2020resttestgen}. Hence, there is a clear need to enhance these testing techniques to overcome these limitations and ensure more comprehensive and realistic testing of RESTful APIs. Notably, our proposed approach ({\method}) focuses on reducing the costs associated with unsuccessful HTTP calls, further optimizing the testing process.

One of the most relevant works is the study conducted by Mirabella et al.~\cite{mirabella2021deep}. While their work shares a common goal of leveraging ML techniques for API testing, there are notable differences between their approach and ours. Mirabella et al. focused on predicting the validity of test inputs by employing a deep learning-based approach to predict whether test inputs satisfy all input constraints. In contrast, we focus on predicting the success or failure of API requests generated by testing tools, considering test inputs as a whole, and encompassing the entire request-response cycle. By predicting the status codes associated with API responses, our approach reduces the number of requests to be executed while maintaining the same effectiveness of the API functionality under test.

\section{Conclusion}\label{conclusion}
This work focused on devising a solution for testing a real-world evolving application, i.e., GURI from the Cancer Registry of Norway (CRN). We presented the EvoMaster extension {\method}, which utilizes machine learning to reduce the cost of testing GURI. We evaluated the cost-effectiveness of {\method} using GURI's ten versions under three environments. The results show that {\method} can significantly reduce testing cost (i.e., $\approx$31\%), meanwhile achieving rule coverage similar to the default EvoMaster. In the future, we plan to propose domain-specific test generation methods such as rule coverage to improve the effectiveness further. In addition, we plan to integrate our solution with testing tools other than EvoMaster. Finally, we want to test other software systems from CRN.  

\section*{Acknowledgment}
This work is supported by the "AI-Powered Testing Infrastructure for Cancer Registry System" project (No. \#309642) funded by the Research Council of Norway. The Norwegian Ministry of Education and Research supports Erblin Isaku's Ph.D.   
The experiment was conducted on the Experimental Infrastructure for Exploration of Exascale Computing (eX3), which is financially supported by RCN under contract 270053. 

\bibliographystyle{IEEEtran}
\bibliography{refs}

\end{document}